\documentclass[10pt,conference]{IEEEtran}
\IEEEoverridecommandlockouts
\usepackage{cite}
\usepackage[numbers]{natbib}
\usepackage{amsmath,amssymb,amsfonts}
\usepackage{algorithmic}
\usepackage{graphicx}
\usepackage{booktabs}
\usepackage{multirow}
\usepackage{textcomp}
\usepackage{xcolor}
\usepackage[most]{tcolorbox}
\def\BibTeX{{\rm B\kern-.05em{\sc i\kern-.025em b}\kern-.08em
    T\kern-.1667em\lower.7ex\hbox{E}\kern-.125emX}}
\begin{document}

\title{Automating Method Naming with Context-Aware Prompt-Tuning
\thanks{$^{\ast}$Li Yang is the corresponding author.}
}

\makeatletter
\newcommand{\linebreakand}{%
  \end{@IEEEauthorhalign}
  \hfill\mbox{}\par
  \mbox{}\hfill\begin{@IEEEauthorhalign}
}
\makeatother


\author{
\IEEEauthorblockN{Jie Zhu}
\IEEEauthorblockA{\textit{Institute of Software, Chinese} \\
\textit{Academy of Sciences, University of}\\
\textit{Chinese Academy of Sciences} \\
Beijing, China \\
zhujie212@mails.ucas.ac.cn}
\and
\IEEEauthorblockN{Lingwei Li}
\IEEEauthorblockA{\textit{Institute of Software, Chinese} \\
\textit{Academy of Sciences, University of}\\
\textit{Chinese Academy of Sciences} \\
Beijing, China \\
lilingwei20@mails.ucas.ac.cn}
\and
\IEEEauthorblockN{Li Yang*}
\IEEEauthorblockA{\textit{Institute of Software, Chinese} \\
\textit{Academy of Sciences}\\
Beijing, China \\
yangli2017@iscas.ac.cn}
\linebreakand 
\IEEEauthorblockN{Xiaoxiao Ma}
\IEEEauthorblockA{\textit{Institute of Software, Chinese} \\
\textit{Academy of Sciences}\\
Beijing, China \\
xiaoxiao@iscas.ac.cn}
\and
\IEEEauthorblockN{Chun Zuo}
\IEEEauthorblockA{\textit{Sinosoft Company Limited} \\
Beijing, China \\
zuochun@sinosoft.com.cn}
}

\maketitle

\begin{abstract}
Method names are crucial to program comprehension and maintenance. Recently, many approaches have been proposed to automatically recommend method names and detect inconsistent names. Despite promising, their results are still sub-optimal considering the three following drawbacks: 1) These models are mostly trained from scratch, learning two different objectives simultaneously. The misalignment between two objectives will negatively affect training efficiency and model performance. 2) The enclosing class context is not fully exploited, making it difficult to learn the abstract functionality of the method. 3) Current method name consistency checking methods follow a generate-then-compare process, which restricts the accuracy as they highly rely on the quality of generated names and face difficulty measuring the semantic consistency.

In this paper, we propose an approach named AUMENA to AUtomate MEthod NAming tasks with context-aware prompt-tuning. Unlike existing deep learning based approaches, our model first learns the contextualized representation(i.e., class attributes) of programming language and natural language through the pre-training model, then fully exploits the capacity and knowledge of large language model with prompt-tuning to precisely detect inconsistent method names and recommend more accurate names. To better identify semantically consistent names, we model the method name consistency checking task as a two-class classification problem, avoiding the limitation of previous generate-then-compare consistency checking approaches. Experiment results reflect that AUMENA scores 68.6\%, 72.0\%, 73.6\%, 84.7\% on four datasets of method name recommendation, surpassing the state-of-the-art baseline by \underline{8.5\%}, \underline{18.4\%}, \underline{11.0\%}, \underline{12.0\%}, respectively. And our approach scores 80.8\% accuracy on method name consistency checking, reaching an \underline{5.5\%} outperformance. All data and trained models are publicly available. 

\end{abstract}


\section{Introduction}
Program comprehension is the foundation of software evolution and maintenance \cite{DBLP:conf/iwpc/RajlichW02}. Developers need to understand programs written by themselves or others before making any modifications. Among all the factors affecting the program's understandability, method names play a significant role as they are brief summaries of source code and could indicate the developer's purpose \cite{DBLP:journals/tse/KoMCA06}. Recent studies show that some programmers even take notes of crucial method names to help them figure out the application procedure, which further demonstrates its significance in program comprehension \cite{DBLP:conf/icse/RoehmTKM12}.

However, method names could also be confusing, making programs even harder to understand \cite{DBLP:journals/tse/ArnaoudovaEPOAG14, DBLP:conf/iwpc/LawrieMFB06} and more error-prone  \cite{DBLP:conf/wcre/ButlerWYS99a}. To improve the readability and maintainability of software, developers need to rename methods when their original names are of poor quality or their implementation logic has been changed \cite{DBLP:conf/kbse/PerumaMDN18}. Nevertheless, naming is a time-consuming and non-trivial task for developers \cite{DBLP:conf/icse/Liu0BKKKKT19}. Poor method naming still widely occurs in many projects for a variety of reasons including insufficient communication and lack of knowledge about the project \cite{DBLP:conf/csmr/ArnaoudovaPAG13, DBLP:conf/icsm/HigoK12, DBLP:journals/ese/KimK16, DBLP:conf/sigsoft/Wang0LM21}. Besides, according to a recent study, developers often discussed renaming in pull request reviewing activities, aiming at a large range of objectives from fixing typos to keeping naming consistency and better reflecting code responsibility \cite{DBLP:journals/tosem/PantiuchinaZSPO20}. Therefore, automatically detecting inconsistent method names and recommending better method names are especially practical for real-world projects and could boost the researches of related areas such as Code Review \cite{DBLP:journals/tse/ScalabrinoBVLPO21, DBLP:conf/icse/TufanoMMPPB22}, Code Smell \cite{DBLP:conf/iwpc/HanT0CL21}, and Code Refactoring \cite{DBLP:conf/kbse/PerumaMDN18, DBLP:journals/tosem/PantiuchinaZSPO20}.

Recently, various automatic approaches have been proposed on the two tasks of method naming: 1) Method name Consistency Checking(MCC), and 2) Method Name Recommendation(MNR). With the idea that similar methods should have similar names, early studies \cite{DBLP:conf/icse/Liu0BKKKKT19, DBLP:conf/apsec/YonaiHK19, DBLP:journals/pacmpl/AlonZLY19} such as Code2Vec \cite{DBLP:journals/pacmpl/AlonZLY19} and Mercem \cite{DBLP:conf/apsec/YonaiHK19} mainly employ information retrieval(IR) techniques to recommend the names of similarly implemented methods. However, these IR-based methods fail to generate new names which have never been seen before. Code2Seq \cite{DBLP:conf/iclr/AlonBLY19} further improves IR-based Code2Vec \cite{DBLP:journals/pacmpl/AlonZLY19} by importing seq2seq models to recommend method names. Nguyen et al. \cite{DBLP:conf/icse/NguyenPLN20} proposed MNire which utilized the enclosing class name to enhance method name suggestion. Li et al. \cite{DBLP:conf/icse/Li0N21} developed DeepName and extended the model input context to surrounding and interacting methods. Wang et al. \cite{DBLP:conf/sigsoft/Wang0LM21} and Liu et al. \cite{DBLP:conf/icse/LiuLFLHJ22} further analyzed and combined contexts from different levels, achieving state-of-the-art results on MNR. For the task of method name consistency checking, the state-of-the-art approach Cognac \cite{DBLP:conf/sigsoft/Wang0LM21} follows the same strategy of MNire \cite{DBLP:conf/icse/NguyenPLN20}, which checks method name consistency by generating a method name first and then computing the lexical similarity between the newly generated name and current name. If the calculated similarity is lower than the selected threshold, the current name will be labeled as inconsistent.

Despite promising results, these previous method naming approaches have three principal limitations. \textbf{1) All models of deep learning based methods are trained from scratch, learning two distinct objectives simultaneously}: one is to learn the semantic representation of programming language and natural language, and another is about learning the relationship between the method name and its implementation to check consistency and recommend better names. The misalignment between the two optimizing objectives decreases the efficiency of training and thus leads to sub-optimal results. \textbf{2) Most recent method name consistency checking approaches, including MNire \cite{DBLP:conf/icse/NguyenPLN20} and Cognac \cite{DBLP:conf/sigsoft/Wang0LM21}, follow a generate-then-compare strategy to detect inconsistent method names, facing difficulty measuring the semantic consistency.} These MCC models determine whether a method name is consistent by calculating the lexical similarity between the current method name and newly generated name. If the calculated lexical similarity is lower than the selected threshold, the original method name will be labeled as inconsistent, otherwise not. However, method names with completely different sub-tokens could be semantically similar while names with low lexical similarity could have totally different meanings. For example, method names "\textit{create\_form\_data}" and "\textit{delete\_form\_data}" are similar in literal, but convey the opposite meanings. And previous generate-then-compare MCC methods will incorrectly predict this example to be consistent as they only consider the lexical similarity. \textbf{3) The context in the enclosing class is not fully exploited.} Previous studies \cite{DBLP:conf/icse/NguyenPLN20, DBLP:conf/icse/Li0N21, DBLP:conf/sigsoft/Wang0LM21} focus on the class name and the sibling methods without considering class attributes.


To mitigate the above issues, we propose AUMENA, a method naming automation approach. First, we adopt the “pre-train, prompt, and predict” paradigm to detect inconsistent method names and generate high-quality names. Specifically, with the pre-trained model that has learned the optimal contextualized representation of code tokens and natural texts in advance, AUMENA could concentrate more on the downstream tasks of method naming. Besides, the CodeT5 \cite{DBLP:conf/emnlp/0034WJH21} pre-trained model adopted in our approach has been specially pre-trained using the identifier tagging and masked identifier prediction tasks. These two elaborately designed tasks highlight the importance of identifiers in source code, which perfectly fit our model and tasks since identifiers play a vital role in method naming. In addition, we involve prompt tuning to bridge the gap between pre-training and tuning on downstream tasks, which contributes to fully exploiting the knowledge and capacity of the pre-trained model. Second, AUMENA models the method name consistency checking task as a 2-class classification problem. Given the method name and the method contexts, the prompt-based binary classification model will directly predict whether the method name is consistent, avoiding the disadvantages of calculating lexical similarity. The major challenge in building the classification model lies in how to collect and construct sufficient inconsistent method names for training, as they should be related but not consistent with the implementation. To cope with this challenge, we propose a novel hard negative sampling method to generate sufficient high-quality negative training samples for the classification model. Finally, AUMENA involves class attributes to enrich the enclosing context, which proves to be effective in method name recommendation. 

To evaluate the effectiveness of AUMENA on method name recommendation task, we conducted experiments following Cognac \cite{DBLP:conf/sigsoft/Wang0LM21} and GTNM \cite{DBLP:conf/icse/LiuLFLHJ22} on four widely-adopted datasets, including \textit{Java-small}, \textit{Java-med} and \textit{Java-large} from Alon et al. \cite{DBLP:conf/iclr/AlonBLY19} and another one built by Nguyen et al. \cite{DBLP:conf/icse/NguyenPLN20}. Experimental results reflect that AUMENA outperforms all state-of-the-art approaches by a large margin (e.g., AUMENA surpass existing techniques by 8.5\%, 18.4\%, 11.0\%, and 12.0\%, respectively, on F-score of four datasets). For the method name consistency checking task, our approach also achieves better performance on the test set collected by Liu et al. \cite{DBLP:conf/icse/Liu0BKKKKT19}, improving from 76.6\% to 80.8\% on the overall accuracy compared with the state-of-the-art Cognac \cite{DBLP:conf/sigsoft/Wang0LM21}.

The main contributions of our approach are outlined as follows:
\begin{itemize}
  \item To the best of our knowledge, we are the first to leverage prompt-tuning on method naming, which exploits the potential of pre-trained models by filling the gap between the pre-training tasks and downstream naming tasks. 
  \item We propose a prompt-based binary classification approach to detect inconsistent names, which is capable of measuring the semantic consistency between method name and its implementation.
  \item Experiment results on five widely-used datasets show that AUMENA performs better than all state-of-the-art approaches by a large margin on both the method name recommendation and method name consistency checking tasks. And the quality of AUMENA-generated method names is similar or even higher than human-written ones.
\end{itemize}

All trained models and data are publicly available \footnote{https://figshare.com/s/0382ba979d970b4c2b23}.

\begin{figure*}[ht]
  \centering
  \includegraphics[width=0.9\textwidth]{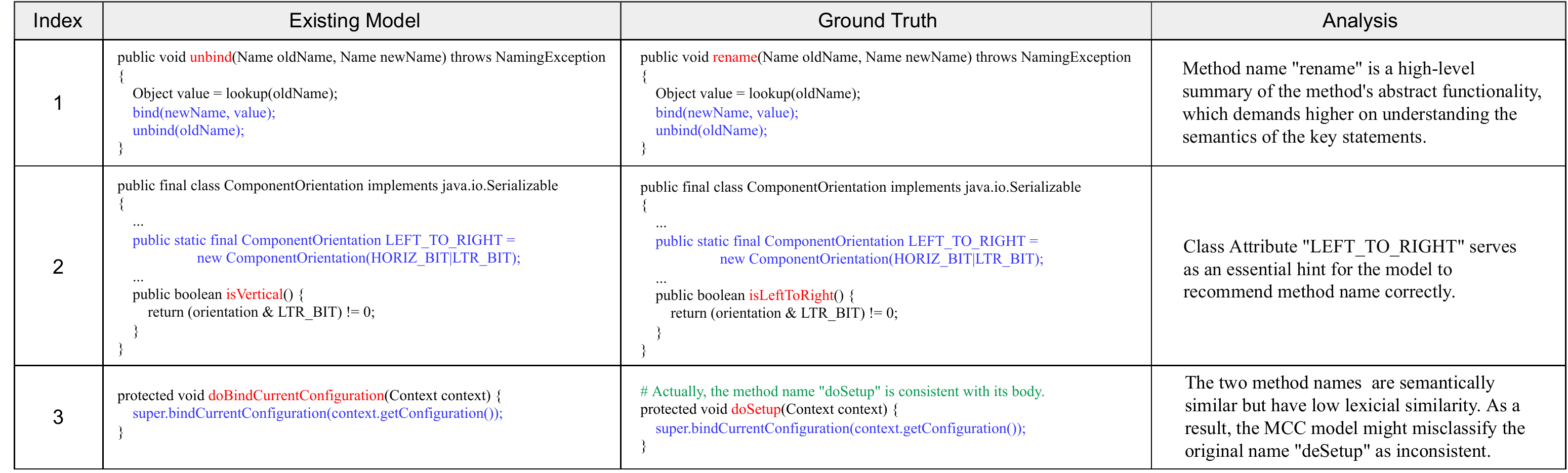}
  \caption{Motivating Examples}
  \label{motexamples}
\end{figure*}

\section{Motivating Examples}


Figure \ref{motexamples} shows three motivating examples in our dataset. Example 1 and 2 are on the method name recommendation task and Example 3 is about detecting inconsistent names.

In Example 1, it shows that existing models suffer from a discrepancy in that method names sometimes could not be extracted from tokens of program entities directly. It can be observed that the method aims to change the name of the object "value" with "bind" and "unbind" operations. However, "unbind" recomended by existing MNR model is just one of the two operations to activate the "oldName" variable. And the whole statement of the two operations "bind(newName, value); unbind(oldName);" is too long to be a method name. e. In that case, we intend to improve it by leveraging pre-trained models since they are able to understand the semantics of the key statement and summarize it as a phrase "rename".

Example 2 manifests the significance of introducing class attributes in Method Name Recommendation. From the example, we can learn that the input method is not sufficient to present the entire function of the code. And that's why the prior study failed to recommend this name. However, we observed that the context of class attributes could help to fill up the lack. In this situation, the statement above "public static final ComponentOrientation LEFT\_TO\_RIGHT = new ComponentOrientation(HORIZ\_BIT$|$LTR\_BIT);" provides a target variable the method will work on. We improve the framework of method name recommendation following this message.

Example 3 reflects the limitation of previous generate-then-compare method name consistency checking approaches. They detect inconsistent names by calculating the lexical similarity between the original method name and generated name. According to this case, the original method name "doSetup" is consistent with its body. However, existing consistency checking model marks it as inconsistent in that the lexical similarity between "doSetup" and the recommended name "doBindCurrentConfiguration" is relatively small, even though they are semantically similar. This further demonstrates the disadvantages of previous generate-then-compare methods. And we could develop a classification-based consistency checking approach to avoid the drawback.

\section{Related Work}

\subsection{Method Name Recommendation}
Recently, many automatic approaches have been proposed to ease the burden of method naming. By constructing two vector spaces of names and bodies to compute similarities, Liu et al. \cite{DBLP:conf/icse/Liu0BKKKKT19} built a deep learning model to recommend method names by retrieving names from most similar methods. Yonai et al. \cite{DBLP:conf/apsec/YonaiHK19} proposed Mercem, which suggests method names based on the method embedding obtained from the caller-callee relationship. Code2Vec \cite{DBLP:journals/pacmpl/AlonZLY19} exploits structural properties within a code snippet by introducing the abstract syntax tree (AST) and then retrieves the most semantically similar method names. However, the IR-based methods above face difficulty generating new names have never seen before, making them harder to be applied in practice. To tackle the above issue, Code2Seq \cite{DBLP:conf/iclr/AlonBLY19} further improves Code2Vec \cite{DBLP:journals/pacmpl/AlonZLY19} by using the seq2seq model to generate method names. Allamanis et al. \cite{DBLP:conf/sigsoft/AllamanisBBS15} developed a neural probabilistic language model for learning the embedding of source code tokens to generate method names. Later they improved their model by utilizing a neural convolutional attentional network, which requires no hard-coded features to extract local time-invariant and long-range topical attention features \cite{DBLP:conf/icml/AllamanisPS16}. Xu et al. \cite{DBLP:conf/pepm/XuZWCGX19} used the hierarchical neural network to suggest method names with block structures. The limitation of these works above is that only the context within the target itself is leveraged. To overcome the restriction, Nguyen et al. \cite{DBLP:conf/icse/NguyenPLN20} proposed MNire which utilized program entities including the enclosing class name to recommend method names. Following the idea of MNire, Li et al. \cite{DBLP:conf/icse/Li0N21} developed DeepName, which extends the model input context to surrounding and interacting methods, such as caller, callee, and sibling methods. Wang et al. \cite{DBLP:conf/sigsoft/Wang0LM21} conducted an empirical study on the relationship between method name and diverse specific contexts and proposed Cognac, a method name recommendation model based on pointer-generator network \cite{DBLP:conf/acl/SeeLM17}. Ge et al. \cite{DBLP:conf/iwpc/GeK21} built a two-phase keywords guided approach which decomposes the method naming task into keywords extraction and method name generation tasks. Qu et al. \cite{DBLP:conf/wcre/QuHZCY22} proposed SGMNG, a structure-guided method name recommendation approach combining semantic and structural features with a graph neural network. Liu et al. \cite{DBLP:conf/icse/LiuLFLHJ22} built a Transformer-based neural model for method name suggestion, considering contexts of different levels simultaneously.

\subsection{Method Name Consistency Checking}
The method name consistency checking task is to determine whether a given method name is consistent with its implementation, which could also be viewed as the first step prior to method name recommendation. Høst and Østvold \cite{DBLP:conf/ecoop/HostO09} studied the method naming conventions and utilized the convention to extract static rules to debug method names. Kim et al. \cite{DBLP:journals/ese/KimK16} relied on the code dictionary built from API documents to detect inconsistent identifiers including method names. With the intuition that the methods with similar implementation should have similar names, Liu et al. \cite{DBLP:conf/icse/Liu0BKKKKT19} detected inconsistent method names by calculating the similarity between the set of names with similar method name embedding and the set of names with similar method implementation embedding. However, methods sharing similar implementations could still have different names as they might exist in different classes and projects \cite{DBLP:conf/icse/LiuLFLHJ22}. To overcome this shortcoming, in recent years, MNire \cite{DBLP:conf/icse/NguyenPLN20} and DeepName \cite{DBLP:conf/icse/Li0N21} utilized multiple contexts to enhance method name consistency checking. The state-of-the-art approach Cognac \cite{DBLP:conf/sigsoft/Wang0LM21} further explored different contexts and followed the same strategy of MNire \cite{DBLP:conf/icse/NguyenPLN20}, which determines whether a method name is consistent by calculating the lexical similarity between the given method name and the name generated by the method name recommendation model.

\subsection{Prompt Tuning}
In recent years, pre-trained models for source code (e.g., CodeT5 \cite{DBLP:conf/emnlp/0034WJH21}, CodeBERT \cite{DBLP:conf/emnlp/FengGTDFGS0LJZ20}, GraphCodeBERT \cite{DBLP:conf/iclr/GuoRLFT0ZDSFTDC21}) have achieved remarkable success in various code understanding and generation tasks. Most existing researches follow a general "pre-train, fine-tune" paradigm. Specifically, the pre-trained models are first pre-trained on a large corpus of source code using the Masked Language Modeling (MLM) objective and then fine-tuned on a smaller dataset to adapt themselves to the downstream tasks. However, the inputs and optimizing objectives differ during the pre-training and fine-tuning phases. This misalignment makes the knowledge and capacity of pre-trained models hard to be fully exploited and thus limits their performance on the downstream tasks. 

To fill the gap between pre-training and fine-tuning, there are two main research directions: one direction is to redesign the pre-training tasks to make them more similar to the downstream tasks \cite{DBLP:conf/sigsoft/LiLGDJJMGSFS22}, and another is to reformulate the downstream tasks into a fill-in-blank form with prompt which resembles the MLM task in pre-training \cite{DBLP:conf/sigsoft/WangYGP0L22}. Prompt-tuning follows the second direction to mitigate the issues of fine-tuning by providing a prompt template such as "The method name is [\textit{MASK}]" at the end of the input. The training objective of prompt-tuning is to predict the masked token [\textit{MASK}]. The model could further decide whether the method name is consistent using a verbalizer which maps the predicted word to the class. And this goes the same for method name recommendation. Through the improvement of converting the downstream task into the fill-in-blank task, the pre-trained models conduct the downstream task in the same way during pre-training, which assists in bridging the gap and consequently boosts the performance.

\begin{figure*}[ht]
  \centering
\includegraphics[width=0.9\textwidth]{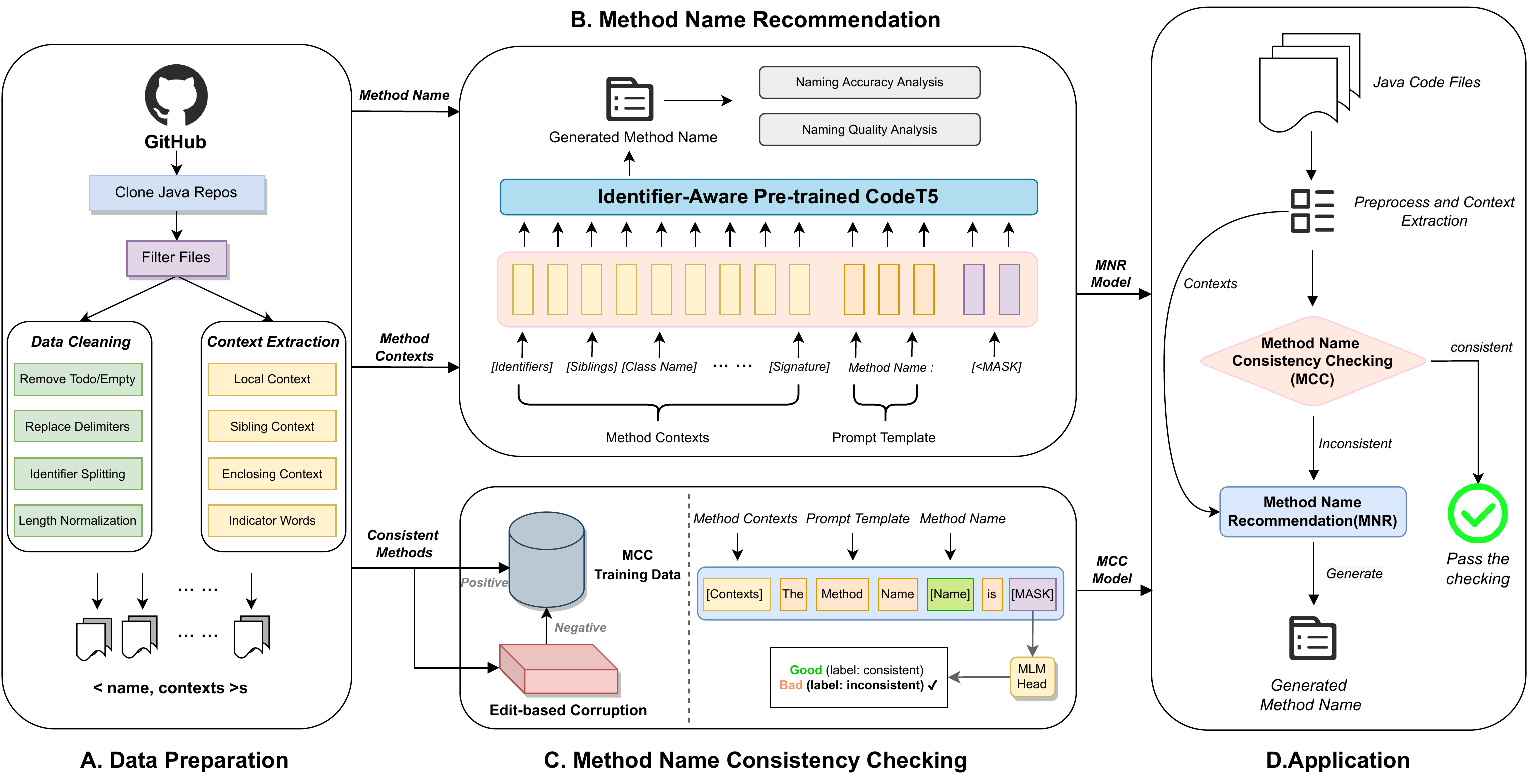}
\caption{Overview of AUMENA}
\label{overview}
\end{figure*}

\section{Approach}

AUMENA mainly focuses on two different but related method naming tasks with prompt-tuning: 1) Method Name Consistency Checking (MCC) and 2) Method Name Recommendation (MNR). We model the MCC task as a 2-class classification problem and the MNR task as a bi-model summarization problem. The overall architecture of our prompt-tuning based method naming approach is depicted in Figure \ref{overview}. We adopt a novel “pre-train, prompt, and predict” paradigm to detect inconsistent method names and generate high-quality candidates. Specifically, for the method name consistency checking task, we propose a novel hard negative sampling method to generate sufficient related but inconsistent method names for training. Afterwards, with the training data collected from the corpus and the sampling approach, we train a 2-class classification model with prompt-tuning to detect inconsistency between method names and their implementations. For the method name recommendation task, AUMENA extracts different contexts from the target method and integrates them with the prompt template we defined as a sequence to feed into the prompt-tuned CodeT5. Compared with the existing approaches based on deep learning, our CodeT5-based model could concentrate more on the task of method naming with the identifier-aware pre-training. Besides, prompt-tuning also helps the pre-trained model better adapt to the two downstream naming tasks by reformulating them into cloze-style fill-in-blank problems that are closer to the pre-training tasks, thus achieving better performance.

\subsection{Data Preparation}
Following the settings of most recent studies \cite{DBLP:conf/icse/LiuLFLHJ22, DBLP:conf/sigsoft/Wang0LM21}, we extract method contexts from different levels and divide the names of these program entities into sequences of sub-tokens using javalang \footnote{https://github.com/c2nes/javalang} and spiral \footnote {https://github.com/casics/spiral}. Specifically, the contexts employed by our approach mainly consist of two parts: (1) local implementation, such as the identifier names and method signature; (2) enclosing class, including class name, siblings, and class attributes. The contexts from identifier names and enclosing classes could provide information on the method's function, while the contexts from siblings and other contextual method names help the model adapt to the naming conventions within specific projects. 
Finally, we convert all input contexts into lowercase sub-token sequences using spiral, as the identifier names are usually compound words.

Considering the length of the sub-token sequence of each context might vary among different examples, we normalize the input data by restricting the length of each context. For example, some classes might have lots of attributes and methods, which might significantly increase the length of these contexts, thus affecting the overall performance as the model might ignore other shorter contexts. Consequently, we limit the number of sibling methods and class attributes to ten. And the maximum length of total input sequences is set to 512.

\subsection{Method Name Recommendation}
\label{sec:mnr}

In method name recommendation, we employ similar contexts with method name consistency, except that we further involve class contexts as they are not available in the MCC dataset of Liu \cite{DBLP:conf/icse/Liu0BKKKKT19}. Compared to previous studies on MNR, we ease the model’s burden of learning the representation of programming and natural language by prompt-tuning the identifier-aware pre-trained model CodeT5 \cite{DBLP:conf/emnlp/0034WJH21}, which could provide a better initialization and help the model concentrate more on learning the relationship between method names and their contexts. Specifically, AUMENA will concatenate all the method contexts with indicator words and then append a natural language prompt template text to guide the model to output method names. For evaluation, we utilize an automated method name quality accessing tool \cite{DBLP:conf/iwpc/AlsuhaibaniNDCM22} in addition to the quantitative metrics on accuracy.

To train the method name recommendation model, we use a cross-entropy loss here to measure the difference between the prediction and the truth. Equation \ref{equ:mnr-loss} accumulates the sum of each sentence predicting errors (i.e., the predicted words $\hat{y}_{ij}^D$ and the manual ones $y_{ij}^D$ in dataset $D$). And the model is trained to reduce the loss as follows.
\begin{equation}
\begin{aligned}
  &LOSS_{MNR} = -\frac{1}{|D|} \sum_{i=1}^{|D|} \sum_{j=1}^{|Y|} y_{ij}^D log(\hat{y}_{ij}^D)
\end{aligned}
\label{equ:mnr-loss}
\end{equation}

\subsection{Method Name Consistency Checking}
\label{sec:hard_neg_section}

\subsubsection{Background}
Similar to prior work \cite{DBLP:conf/icse/Liu0BKKKKT19, DBLP:conf/icse/NguyenPLN20, DBLP:conf/sigsoft/Wang0LM21} on Method name Consistency Checking(MCC), AUMENA takes a method with its name and contexts as input and predicts whether the given method name is consistent with its implementation. However, previous state-of-the-art MCC approaches \cite{DBLP:conf/icse/NguyenPLN20, DBLP:conf/sigsoft/Wang0LM21} follow a generate-then-compare strategy. They detect inconsistent method names by generating a new method name given its implementation first and then calculating the lexical similarity between the generated name and the original name. If the similarity calculated is lower than the selected threshold, the original method name will be regarded as inconsistent. However, the limitation of these approaches is that the lexical similarity could not reflect the semantics of words. Two method names with totally different words could still have similar meanings. To address the above issues, we propose to model this task as a binary classification problem. The prompt-based 2-class classification model takes the method name with its implementation together and directly predicts whether the method name is consistent. And the main challenge of building this classification model lies in collecting sufficient training data, especially the inconsistent examples. 

\subsubsection{Hard Negative Sampling to Build MCC Training Data}
To acquire adequate training data for classification, we regard the methods from the datasets of method name recommendation as consistent examples, as they have been stable for a long time \cite{DBLP:conf/icse/Li0N21}. Nevertheless, inconsistent methods are somewhat difficult to collect, and there are no large-scale open datasets currently. To the best of our knowledge, the only inconsistent naming dataset is collected by Liu \cite{DBLP:conf/icse/Liu0BKKKKT19}. However, it only contains 2805 inconsistent examples, which are totally inadequate for training. To cope with this challenge, we propose a novel hard negative sampling method to build sufficient inconsistent naming examples from consistent ones. And the main difficulty here is about how to generate related but inconsistent names. For example, if we randomly choose names from other methods in the corpus as inconsistent examples, the classification model could easily identify them as they are totally unrelated to the method implementation. However, inconsistent method names are usually closely related to their contexts, and most of them could be transformed into consistent ones by several edits. Motivated by this observation, we decided to take an edit-based method to generate fake inconsistent names from consistent ones. Specifically, for a given consistent method name, we randomly corrupt the sub-tokens of names with the probability estimated from Liu’s datasets, and the operations conducted on each token include “remain the same”, “add”, “delete” and “replace”. Finally, we generate 4M inconsistent names and combine them with positive examples to form the training dataset of the method name consistency checking model.

\subsubsection{Prompt-based 2-class Classification Model}
After acquiring adequate training data, we combine the method name and contexts with the prompt template to optimize our classification model. Given a training example $x$ with its name $n$ and contexts $c$, the prompt-based classification model will predict the probability distribution among the label words. And a verbalizer $v$ will map the predicted label word to the corresponding class, which is the final output of the classification model. And the loss of MCC is as follows:

\begin{equation}
\begin{aligned}
  &LOSS_{MCC} = -\frac{1}{|D|} \sum_{i=1}^{|D|} \delta_{i}^D log(\hat{\delta}_{i}^D)
\end{aligned}
\label{equ:mcc-loss}
\end{equation}

A cross-entropy loss is employed to measure the deviation between the prediction and the truth. Equation \ref{equ:mcc-loss} accumulates the sum of each true-or-false classification entropy (i.e., the predicted $\hat{\delta}_{i}^D$ and the golden one $\delta_{i}^D$ in dataset $D$). 

\subsection{Application}

After finishing the training of the MNR and MCC models, AUMENA automates the method naming tasks in three steps. First, we extract the necessary contexts, including identifiers, signature, and siblings, etc., from given java source files and send them into the MCC model. In the second step, the MCC model will determine whether the given method name is consistent with its implementation with prompt-based 2-class classification. If the predicted output is positive(consistent), the process will end as the given method passes the consistency check. If negative, the MNR model will take the contexts extracted from the first step to generate new candidate names. And we call this process consisted of three steps as the method naming automation problem.

\section{Experimental Design}

\subsection{Research Questions}

To evaluate our proposed AUMENA approach, we conduct experiments to answer the following research questions.

RQ1: How does AUMENA perform on the method name recommendation task against other state-of-the-art methods?

RQ2: How does AUMENA perform on the method name consistency checking task against other baselines?

RQ3: How effective is the design of modeling method name consistency checking task as a prompt-based binary classification problem? 

RQ4: How is the quality of method names generated by AUMENA, compared with human-written ones? 

\begin{table}
  \centering
  \caption{Statistics of the MNR Datasets}
  \label{tab:datasets}
  \begin{tabular}{c|ccc}
    \toprule
    \textbf{Datasets} &\textbf{Train} &\textbf{Validation} &\textbf{Test}\\
    \midrule
    Java-small &643K &31K &45K \\
    Java-median &2,711K &389K &369K \\
    Java-large &13,442K &305K &403K \\
    MNire's &16,580K &3,982K &267K \\
    \bottomrule
  \end{tabular}
\end{table}

\subsection{Datasets}
\label{sec:dataset}

For method name recommendation tasks, we used four different datasets for evaluation following Cognac \cite{DBLP:conf/sigsoft/Wang0LM21} and GTNM \cite{DBLP:conf/icse/LiuLFLHJ22}. As shown in Table \ref{tab:datasets}, three datasets, including \textit{Java-small}, \textit{Java-med}, and \textit{Java-large} from Alon et al. \cite{DBLP:conf/iclr/AlonBLY19}, contain 719K, 3.47M and 14.2M Java method examples, respectively. And another dataset built by Nguyen et al. \cite{DBLP:conf/icse/NguyenPLN20} has 20.8M Java examples. The first three datasets provide concrete method data with the partition of training, validation, and test set, which could be downloaded directly. For the dataset of MNire by Nguyen et al. \cite{DBLP:conf/icse/NguyenPLN20}, we follow the same settings and partition of GTNM \cite{DBLP:conf/icse/LiuLFLHJ22} to preprocess and build the dataset. To avoid data leakage, all method examples are split by projects instead of file-based validation, as there might be similar methods in the same project from which models could learn the coding convention of test data \cite{DBLP:conf/sigsoft/Wang0LM21, DBLP:conf/kbse/JiangLJ19}.

For method name consistency checking tasks, there is no available large training dataset due to the lack of negative samples. Therefore, the dataset we used for our classification model is built from the method name recommendation corpus of GTNM \cite{DBLP:conf/icse/LiuLFLHJ22} using the hard negative sampling approach described in \ref{sec:hard_neg_section}. To compare our approach against other methods, we evaluated AUMENA on the test set collected by Liu et al. \cite{DBLP:conf/icse/Liu0BKKKKT19}, which has been widely adopted for evaluation by previous studies \cite{DBLP:conf/icse/NguyenPLN20, DBLP:conf/icse/Li0N21, DBLP:conf/sigsoft/Wang0LM21}.

\subsection{Metrics}
For method name recommendation, we followed prior works and used traditional Precision, Recall, and F1-score scores to evaluate our model and other baselines. These metrics measure the similarity between the prediction and the ground truth at the word level. Precision calculates the rate of matches to total predictions. Recall is the rate of matches to all samples in truth. And F1-score is the harmonic mean of Precision and Recall. To better evaluate the generation of the entire names, we also employed Exact Match Accuracy to take the order of sub-tokens into consideration, which proves to be more precise according to recent studies \cite{DBLP:conf/sigsoft/Wang0LM21, DBLP:conf/icse/LiuLFLHJ22}.

For method name consistency checking, we used Precision, Recall, and F1-score for classification, too. In addition, Overall Accuracy was employed to measure how many generations, both positive and negative, were correctly classified, referring to golden ones.


\begin{table}
  \centering
  \caption{Training Hyperparameters}
  \label{tab:hyperparameter}
  \begin{tabular}{c|c|c}
    \toprule
     &Hyperparameter &Value\\
    \midrule
    \multirow{5}{55pt}{Method Name Recommendation}
        &Learning rate &5e-5 \\
        &Max input length &512 \\
        &Max output length &16 \\
        &Beam size &10 \\
        &Batch size &16 \\
    \midrule
    \multirow{3}{55pt}{Method Name Consistency Checking}
        &Learning rate &5e-5 \\
        &Max input length &512 \\
        &Batch size &16 \\
    \bottomrule
  \end{tabular}
\end{table}

\subsection{Experiment Settings}

1) Method Name Recommendation. In this part, we leveraged CodeT5 \cite{DBLP:conf/emnlp/0034WJH21} and followed the initialization of its pretraining work. For prompt-tuning, we trained our model in 1,036K steps and evaluated at every 100K step. Some key hyperparameters are shown in Table \ref{tab:hyperparameter}.

2) Method Name Consistency Checking. Similar to Method Name Recommendation, we selected pretrained CodeT5 model to implement consistency checking, and it took 520K steps for the entire prompt-tuning and was evaluated at every 50K step. Also, Table \ref{tab:hyperparameter} presents some key hyperparameters we set in experiments.

3) Experiment Environment. We implemented all training with three NVIDIA GeForce RTX 3090Ti GPU and CUDA 11.7\footnote{https://developer.nvidia.com/cuda-toolkit/}. To sum up, it took 100 hours for Method Name Recommendation and 23 hours for Method Name Consistency Checking on training, respectively.

\section{Method Name Recommendation Task}

To evaluate AUMENA on method name recommendation task, we compare our approach against other baselines on four widely-adopted datasets introduced in \ref{sec:dataset}.

\subsection{Baselines}
We chose the following competing baselines to compare with AUMENA.

\begin{itemize}
    \item \textbf{Code2vec.} Alon et al. \cite{DBLP:journals/pacmpl/AlonZLY19} employ information retrieval(IR) techniques to recommend the names of similarly implemented methods.
    \item \textbf{MNire.} Nguyen et al. \cite{DBLP:conf/icse/NguyenPLN20} use the sub-tokens of class name and method body to generate method names with RNN-based seq2seq model.
    \item \textbf{Cognac.} Wang et al. \cite{DBLP:conf/sigsoft/Wang0LM21} explore not only the local context but also the global context to recommend method names with pointer-generator network.
    \item \textbf{GTNM.} Liu et al. \cite{DBLP:conf/icse/LiuLFLHJ22} consider contexts from different levels to generate method names using a transformer-based seq2seq model.
\end{itemize}

\begin{table}
  \centering
  \caption{Results of Method Name Recommendation}
  \label{tab:mnr}
  \begin{tabular}{c|c|cccc}
    \toprule
    Dataset &Approach &Precision &Recall &F1-score &EM Acc\\
    \midrule
    \multirow{4}{*}{Java-small}
        &Code2vec &23.4 &22 &21.4 &- \\
        &MNire &44.8 &38.7 &41.5 &15.5 \\
        &Cognac &67.1 &59.7 &63.2 &- \\
    \cmidrule{2-6}
        &AUMENA &\textbf{69.6} &\textbf{67.6} &\textbf{68.6} &\textbf{44.3} \\
    \midrule
    \multirow{4}{*}{Java-med}
        &Code2vec &36.4 &27.9 &31.9 &- \\
        &MNire &62.0 &57.6 &59.7 &36.2 \\
        &Cognac &64.8 &57.3 &60.8 &- \\
    \cmidrule{2-6}
        &AUMENA &\textbf{72.6} &\textbf{71.4} &\textbf{72.0} &\textbf{50.9} \\
    \midrule
    \multirow{4}{*}{Java-large}
        &Code2vec &44.2 &38.3 &41.6 &- \\
        &MNire &63.1 &59.0 &61 &37.4 \\
        &Cognac &71.4 &61.9 &66.3 &- \\
    \cmidrule{2-6}
        &AUMENA &\textbf{74.0} &\textbf{73.2} &\textbf{73.6} &\textbf{55.3} \\
    \midrule
    \multirow{5.5}{*}{MNire's}
        &Code2vec &51.9 &39.8 &45.1 &35.6 \\
        &MNire &66.3 &62.1 &64.2 &42.6 \\
        &Cognac &70.2 &66.8 &68.5 &- \\
        &GTNM &77.0 &74.1 &75.6 &62.0 \\
    \cmidrule{2-6}
        &AUMENA &\textbf{85.1} &\textbf{84.3} &\textbf{84.7} &\textbf{71.0} \\
    \bottomrule
  \end{tabular}
\end{table}

\subsection{Results(RQ1)}

The results in Table \ref{tab:mnr} show that AUMENA totally outperforms all baselines on four method name recommendation datasets. For Java-small, AUMENA achieves 69.6\% on Precision, 67.6\% on Recall, 68.6\% on F1-score, and 44.3\% on Exact Match, which outperforms baselines by at least 3.73\%, 13.2\%, 8.54\%, respectively. And when it comes to Java-med and Java-large, the results go to over 70\% on Precision, Recall, F1-score, and 50\% on Exact Match. Then, we mainly focus on the performance of MNire's, the largest dataset. AUMENA scores highest on every metric and outperforms baselines by 10.5\%, 13.8\%, 12.0\% and 14.5\%. It indicates that our approach is capable of recommending method names more preciously and achieving a stable performance, whatever the situation is. Compared with pointer-generator network based Cognac \cite{DBLP:conf/sigsoft/Wang0LM21}, AUMENA could better handle long context sequences with transformer-based CodeT5 model, since RNN architecture has difficulties remembering long-term dependencies. And our proposed approach also outperformed GTNM \cite{DBLP:conf/icse/LiuLFLHJ22} in that AUMENA could fully exploit the knowledge and capacity of pre-trained model with prompt tuning for better understanding of tokens in programming language and natural language. Thus, AUMENA could concentrate more on learning to generate method name accurately.

\begin{table*}
  \centering
  \caption{Results of Method Name Consistency Checking}
  \label{tab:mcc}
  \begin{tabular}{c|c|c|c|c|c|c|c}
    \toprule
     & &DebugMethodName\cite{DBLP:conf/icse/Liu0BKKKKT19} &MNire\cite{DBLP:conf/icse/NguyenPLN20} &DeepName\cite{DBLP:conf/icse/Li0N21} &Cognac\cite{DBLP:conf/sigsoft/Wang0LM21} &AUMENA* &AUMENA \\
    \midrule
    \multirow{3}{*}{Inconsistent}
        &Precision &56.8 &62.7 &72.3 &68.6 &\textbf{84.4} &81.9 \\
        &Recall &84.5 &93.6 &92.1 &\textbf{97.6} &70.1 &78.9 \\
        &F-score &67.9 &75.1 &\textbf{81.0} &80.6 &76.6 &80.4 \\
    \midrule
    \multirow{3}{*}{Consistent}
        &Precision &72.0 &84.2 &86.4 &\textbf{96.0} &74.4 &79.7 \\
        &Recall &38.2 &56.0	&64.8 &55.6 &\textbf{87.0} &82.6 \\
        &F-score &49.9 &67.3 &74.1 &70.4 &80.2 &\textbf{81.1} \\
    \midrule
    \multicolumn{2}{c|}{Overall Accuracy}
        &60.9 &68.9 &75.8 &76.6 &78.6 &\textbf{80.8} \\
    \bottomrule
  \end{tabular}
\end{table*}

\section{Method Name Consistency Checking Task}

For the method name consistency checking task, we used the widely-used dataset collected by Liu et al. \cite{DBLP:conf/icse/Liu0BKKKKT19} to evaluate the effectiveness of AUMENA.

\subsection{Baselines}
We compare AUMENA with state-of-the-art baselines as follows.

\begin{itemize}
    \item \textbf{DebugMethodName.} Liu et al. \cite{DBLP:conf/icse/Liu0BKKKKT19}  detect inconsistent method names by calculating the similarity between the set of names with similar method name embedding and the set of names with similar implementation embedding. 
    \item \textbf{MNire.} Nguyen et al. \cite{DBLP:conf/icse/NguyenPLN20} first generate a new method name and then compares the current name against it to detect inconsistency. 
    \item \textbf{DeepName.} Li et al. \cite{DBLP:conf/icse/Li0N21} build a two-channel CNN model which takes the representation vectors of method implementation and method name to predict whether it is consistent.
    \item \textbf{Cognac.} Wang et al. \cite{DBLP:conf/sigsoft/Wang0LM21} follow the same strategy of MNire \cite{DBLP:conf/icse/NguyenPLN20}, which computes the lexical similarity between the original name and newly generated name to check consistency.
\end{itemize}

\begin{figure}
\includegraphics[width=\linewidth]{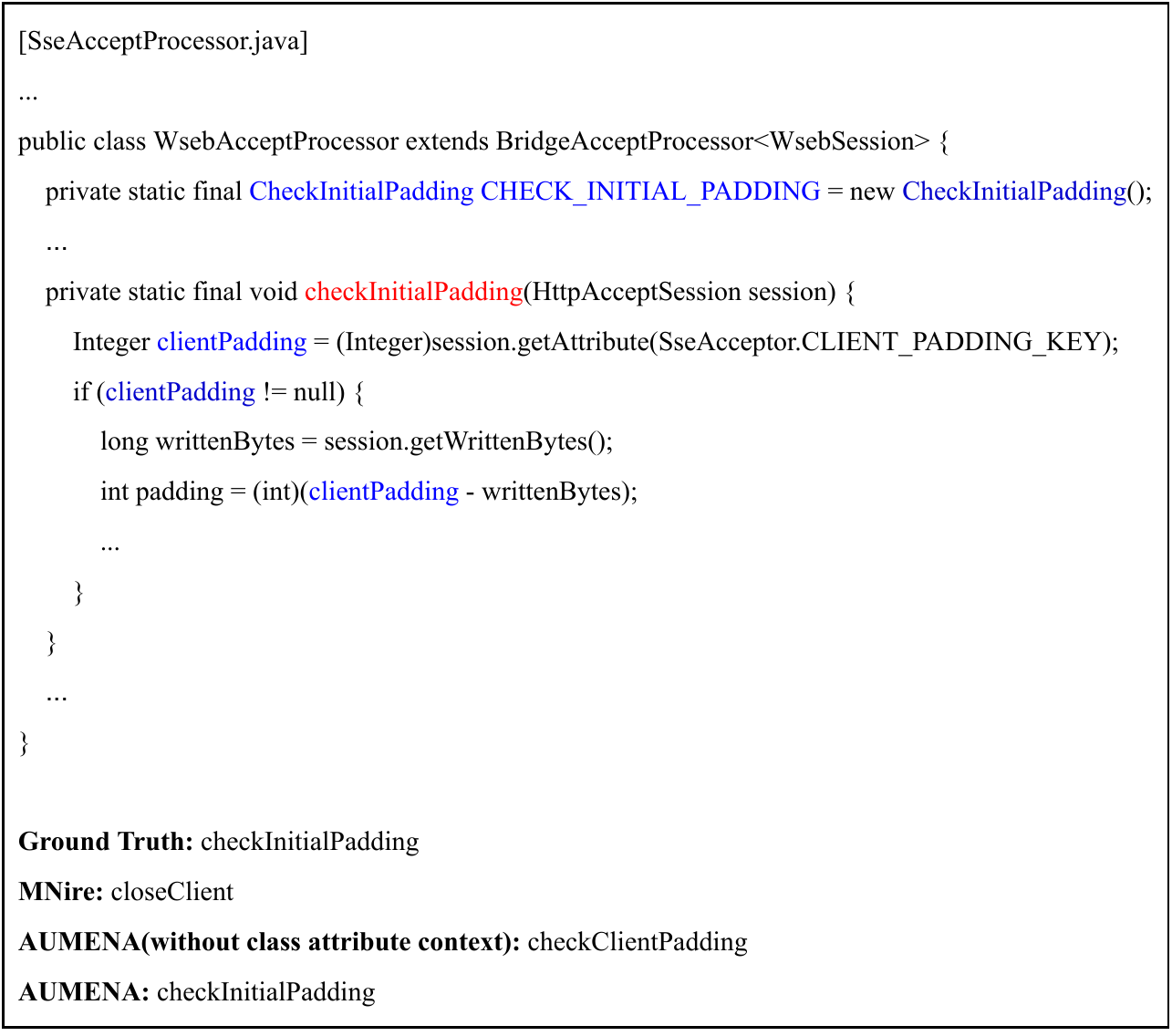}
\caption{MNR Case}
\label{mnrcase}
\end{figure}

\begin{figure*}[ht]
  \centering
\includegraphics[width=1\textwidth]{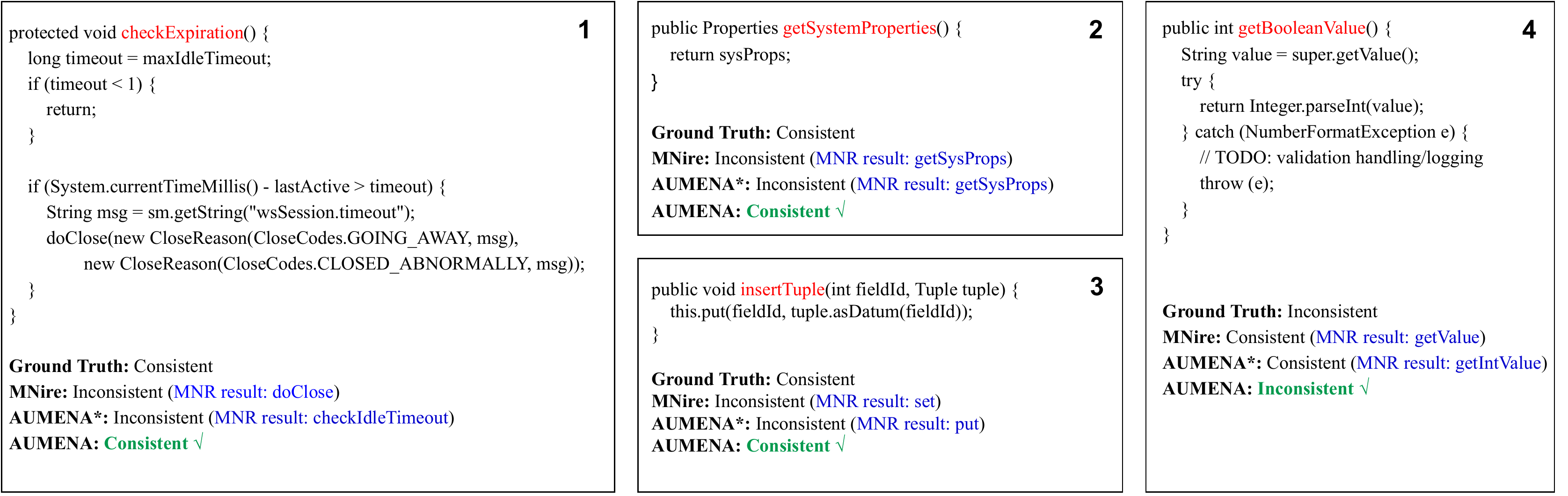}
\caption{Some examples from MCC testset}
\label{mccCase}
\end{figure*}

\subsection{Results(RQ2 \& RQ3)}

Table \ref{tab:mcc} presents the results of Method name Consistency Checking(MCC). To illustrate the superiority of our classification-based approach, we also conducted experiments of AUMENA*, which takes the same strategy as MNire \cite{DBLP:conf/icse/NguyenPLN20} and Cognac \cite{DBLP:conf/sigsoft/Wang0LM21} to detect inconsistency. Specifically, in AUMENA*, we used the trained MNR model to generate a method name first and then compared the current method name against it. If the lexical similarity between them is lower than the selected threshold, AUMENA* will mark the current method name as inconsistent. In this way, we could investigate the contribution of our proposed classification-based approach by comparing results of AUMENA and AUMENA*. 


It can be observed that for inconsistency checking, AUMENA achieves 80.8\% on total accuracy, which leads all baselines by at least 5.5\%. This proves that our model surpasses state-of-the-art on both two opposites for consistency checking, which indicates the effectiveness of AUMENA in method name consistency checking. And we also investigate the contribution of modeling MCC task as a two-class classification problem by comparing the results of AUMENA and AUMENA*. It turns out that our classification-based MCC model(AUMENA) could achieve better performance than AUMENA* on examples from both the consistent and inconsistent ones. And the overall accuracy of AUMENA increases from 78.6\% to 80.8\%, which further demonstrates the superiority of our prompt-based classification approach. In addition, AUMENA* achieves higher overall accuracy compared with  Cognac\cite{DBLP:conf/sigsoft/Wang0LM21} and MNire \cite{DBLP:conf/icse/NguyenPLN20}. This outperformance could also support the argument that our MNR model recommends more accurate method names, since AUMENA* takes the same generate-then-compare strategy of Cognac and MNire. 

\section{Discussion}

\subsection{Qualitative Analysis}

\subsubsection{MNR Case Study}
In this section, we provide a method name recommendation example to demonstrate the context-aware ability of AUMENA. As shown in Figure \ref{mnrcase}, MNire misunderstood the method's functionality and ignored the names of input sibling methods. To further reflect the importance of class attribute context, we train two versions of AUMENA MNR models: one with context from class attribute and another without. In this case, the model with class attribute context gives the correct answer in that the crucial sub-token 'Initial' only appears in the class attribute, which serves as a key clue for the model. This illustrates that AUMENA could consider contexts from different sources to generate more accurate method names.

\subsubsection{MCC Case Study}
Figure \ref{mccCase} presents some examples from the testset of method name consistency checking \cite{DBLP:conf/icse/Liu0BKKKKT19}. Given a method body, MNire \cite{DBLP:conf/icse/NguyenPLN20} and AUMENA* follow the same generate-then-compare strategy to detect inconsistency by calculating the similarity of newly generated name and original name. And the only difference is that MNire takes an RNN-based seq2seq MNR model, while AUMENA* utilizes our CodeT5-based MNR model introduced in \ref{sec:mnr} to recommend method names. In contrast, AUMENA uses our proposed prompt-based classification approach to detect inconsistent method names. Different from AUMENA*, it models the MCC task as a 2-class classification problem, which allows AUMENA to measure the semantic consistency between the method name and method implementation. 

In example 1, the method name "checkExpiration" is obviously consistent with its body. However, both the MNire and AUMENA* predict it to be inconsistent, as the names suggested by their MNR models are totally different from the given name "checkExpiration" on word-level. Even if the implication of "checkIdleTimeout" is similar to the target name, AUMENA fails to predict correctly in that it only decides by calculating the lexical similarity between "checkIdleTimeout" and "checkExpiration". And it goes the same in example 2 and 3. Actually the recommended name "getSysProps" in example 2 conveys exactly the same meaning of "getSystemProperties". But as the lexical similarity between their sub-tokens is relatively small, the original method name "getSystemProperties" is incorrectly marked as inconsistent by AUMENA* and MNire. However, AUMENA could give the correct result as it is capable of measuring the semantic consistency directly. And in example 3, generating the method name "insertTuple" perfectly given its implementation is a challenging task. Consequently, both MNire and AUMENA* fail to give the correct MCC result as they highly rely on the accuracy of generated names. But it is much easier for AUMENA to correctly classify as it is only required to conduct a 2-class classification, instead of generating the whole method name perfectly. In example 4, the method name "getBooleanValue" is inconsistent since the method body is about parsing the input string to an integer value instead of a boolean value. However, both MNire and AUMENA* believe that the given method name is appropriate, as their MNR results share most of the sub-tokens with the target method name. This further demonstrates the limitation of previous generate-then-compare MCC approaches in which different words could express the same meaning while method names sharing similar tokens could be totally different in semantics. 


\subsection{Name Quality Analysis(RQ4)}
Most recent method name recommendation approaches evaluate themselves with metrics of machine learning, such as precision, recall, and EM accuracy. However, method names differ from general natural language sequences because they have specific patterns and rules \cite{DBLP:conf/icse/AlsuhaibaniNDCM21}. As a result, in the evaluation of MNR models, not only the accuracy metrics but also the quality of method names should be taken into consideration. 


\begin{figure}
\includegraphics[width=\linewidth]{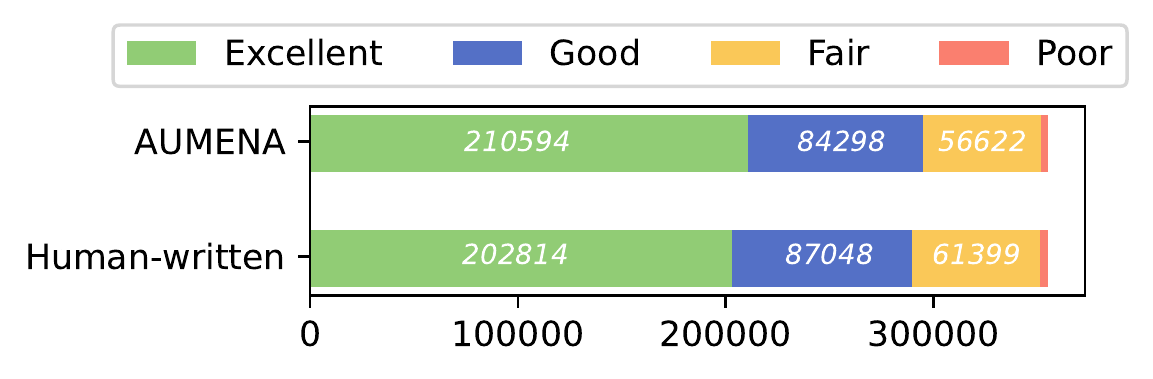}
\caption{Results of Naming Quality Analysis}
\label{quality}
\end{figure}


Specifically, we used the tool developed by Alsuhaibani et al. \cite{DBLP:conf/iwpc/AlsuhaibaniNDCM22} to compare the quality of human-written method names and AUMENA recommended names. The tool implements ten verified method naming standards and produces a score from 0-10 for the given method name based on the ten standards. For example, a score of 10 means the method name follows all the standards and the name will be marked as excellent. Figure \ref{quality} shows the results of naming quality analysis on \textit{java-large} dataset with 355k examples. It could be observed that AUMENA recommends more excellent method names compared with human-written ones. And it turns out that the average score of AUMENA-generated method names is 9.23, which is even higher than the score of human-written method names 9.17. These results of quality analysis illustrate that AUMENA could generate method names with similar or even higher quality compared to human-written ones from the perspective of method naming standards.


\subsection{Name Length Analysis}

We analyze the recommended name length distribution and compare the exact match accuracy of AUMENA with MNire \cite{DBLP:conf/icse/NguyenPLN20} for different name lengths in Figure \ref{fig:length}. Specifically, we apply the Wilcoxon Signed Rank Test (WSRT) \cite{Wilcoxon1945IndividualCB} to figure out whether there is a significant difference between the length distribution of AUMENA-generated method names and original names. The results reflect that there is no significant difference since all the p-values are larger than 0.05. For the exact match accuracy of method names with different lengths, we could find that the EM accuracy decreases as the MNR task with the increase of name length. The reason is that it becomes harder to predict each word correctly when the length of method name increases \cite{DBLP:conf/icse/LiuLFLHJ22}. Still, AUMENA achieves an EM accuracy of 57.12\% for method names of length 5. Compared to the the results of MNire, the EM accuracy curve of AUMENA is much flatter, which illustrates that the accuracy of method names generated by AUMENA is not only better, but also more stable on different lengths.

\begin{figure}
  \centering
  \includegraphics[width=8cm]{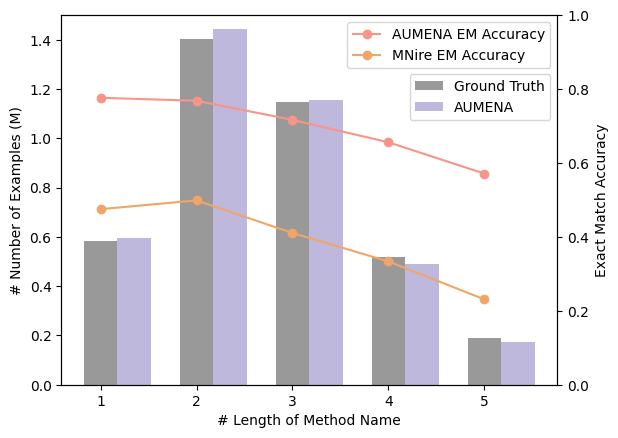}
  \caption{The method name length distribution and the exact match accuracy with different lengths}
  \label{fig:length}
\end{figure}

\subsection{Threats to Validity}

\textbf{Construct validity:} We followed prior studies \cite{DBLP:conf/icse/Liu0BKKKKT19, DBLP:conf/icse/NguyenPLN20, DBLP:conf/icse/Li0N21, DBLP:conf/sigsoft/Wang0LM21, DBLP:conf/icse/LiuLFLHJ22} and constructed experiments on the same datasets. However, since individual knowledge varies, there is no guarantee that all method names are good enough. Also, engineers could keep their own points of view on the same naming work. Therefore, the dataset may still contain sub-optimal method names, and it can affect both the evaluations of method name recommendation and method name consistency checking tasks. Nevertheless, a handful of incorrect samples have little effect on the performance of deep learning based approaches as they learn from the majority instead of the minority \cite{DBLP:conf/sigsoft/Wang0LM21}.

\textbf{Internal validity:} Studies have come up with the argument that the impact of hyperparameters on DL models' effect still remains uncertain \cite{JMLR:v21:20-074, 10.1145/3292500.3330701}. To amplify the performance, we follow Tree-structured Parzen Estimator (TPE) \cite{NIPS2011_86e8f7ab} to optimize all models in this paper. However, we recognize that there may be a few more settings that could cause the same or better results. And that's also a part of our future study.

\textbf{External validity:} To deeply explore the insight of the tasks, we limit our approach to Java projects. Hence, the generalizability for other languages is still unseen in this paper. Future work will move on to explore the performance of AUMENA on other programming languages.


\subsection{Application Scenario and Future Work}
As the experimental results have reflected that AUMENA is more effective than other baselines on both MNR and MCC tasks, it is of significance for us to apply our approach in practice. For example, we could conduct just-in-time name recommendation in real-world software development. And it is also feasible to build a tool to detect inappropriate method names by taking a whole project as input. For the inconsistencies, we could recommend candidate names and organize them as a report for developers as reference. However, in real software engineering projects, inconsistent method names are supposed to be extremely rare. This class-imbalance problem will negatively affect the performance of MCC models in practice. Besides, to what extent the method naming automation tool helps developers remains to be explored further. In the future, we plan to develop a production-ready tool to apply AUMENA in real-world scenarios, for lightening the burden of developers on method naming.

\section{Conclusion}
In this paper, we propose AUMENA, a method naming automation approach based on prompt-tuning. Unlike the prior work, AUMENA develops a prompt-based classification model to detect inconsistent method names, which is capable of measuring the semantic consistency. The novel “pre-train, prompt, and predict” paradigm we adopt helps exploit the potential of pre-trained models by filling the gap between the pre-training tasks and downstream naming tasks. The experimental results show that our approach significantly outperforms other baselines in both the method name recommendation and method name consistency checking tasks. 

\section{ACKNOWLEDGMENTS}
We sincerely appreciate the valuable feedback from the anonymous reviewers. This work was supported by the National Key R\&D Program of China (No.2021YFC3340204), the Alliance of International Science Organizations, (Grant No. ANSO-CR-KP-2022-03), and the Strategy Priority Research Program of Chinese Academy of Sciences (No.XDA20080200).

\bibliographystyle{IEEEtran} 
\bibliography{mybib}

\end{document}